\documentstyle[a4]{article}
\input epsf
\begin{document}
\begin{center}
Charge asymmetry due to the soft pion at high energy 
and its effect on the sea quark\\
Susumu Koretune\\
Department of Physics,Shimane Medical University,Izumo,Shimane,
693-8501,Japan\\
\end{center}

Using the soft pion theorem in the inclusive reactions, we estimate the
soft pion contribution to the structure functions $F_2$ and $g_1$ in the nucleon,
and show that it is an indispensable part in these structure functions in
the small $x$ region. This contribution is asymmetric for the soft $\pi^+$ and $\pi^-$ 
due to pole terms in the soft pion limit, hence it is a remnant of the spontaneous
chiral symmetry breakings. Then we show that the modified strange sea quark where
the soft pion contribution is added to the distribution determined by the ansatz
(strange-sea)=(up-sea + down-sea)/4 satisfies the mean charge sum rule for the sea quarks.  

\section{Introduction} 
A recent measurement of the flavor asymmetry of the light antiquark
in the nucleon by E866/NuSea collaboration \cite{E866} suggests us an importance
of the non-perturbative physics based on the spontaneous chiral symmetry breakings
such as the mesonic models reviewed in \cite{meso}. However all these
models have a valid applicability in the relatively low energy regions.
On the other hand, the modified Gottfried sum rule \cite{Got} has explained 
the NMC deficit in the Gottfried sum \cite{NMC} almost model independently. 
It has shown that the deficit is the reflection of the hadronic vacuum originating 
from the spontaneous chiral symmetry breakings.  The numerical prediction based on 
this sum rule almost exactly agrees with the experimental value from E866/NuSea
collaboration \cite{E866}. In the estimation of this sum rule
we integrate over the cross section for the difference $(\sigma^{K^{+}n}-\sigma^{K^{+}p})$ 
and finds that among the NMC deficit, which we take here $0.07$ for definiteness,
about $40\%$ comes from the region where the momentum of the kaon in the laboratory frame 
is above $4\;$GeV. Further the theoretical basis of this sum rule shows an importance
of the physics not only in the low energy region but also in the high energy region. 
These facts suggest that there may exist a dynamical 
mechanism to produce the flavor asymmetry at medium and high energy which we have
overlooked as yet.
In fact it is shown that the soft pion at high energy can give sizable
contribution to the Gottfried sum\cite{kore99} which may compensate for the
lack of the applicability of the mesonic model in the high energy region.
In this paper, based on this fact we investigate the soft pion contribution
to the strange sea quark distribution and the spin dependent structure
functions.\\
Since the soft pion theorem in the inclusive reaction at high energy
is not well known, let us first explain it briefly. Usually,the soft
pion theorem has been considered to be applicable only in the low
energy regions. However in \cite{sakai}, it has been found that this theorem
can be used in the inclusive reactions at high energy if the Feynman's
scaling hypothesis holds. In the inclusive reaction 
``$\pi + p \to \pi_{s}(k) + anything$'' with the $\pi_{s}$ being the soft pion, 
it states that the differential cross-section
in the center of the mass (CM) frame defined as
\begin{equation}
f(k^3,\vec{k}^{\bot},p^0)=k^0\frac{d\sigma}{d^3k} ,
\end{equation}
where $p^0$ is the CM frame energy, scales as
\begin{equation}
f\sim f^{F}(\frac{k^3}{p^0},\vec{k}^{\bot}) + \frac{g(k^3,\vec{k}^{\bot})}{p^0} .
\end{equation} 
If $g(k^3,\vec{k}^{\bot})$ is not singular at $k^3=0$, we obtain
\begin{equation}
\lim_{p^0\to \infty}f^F(\frac{k^3}{p^0},\vec{k}^{\bot}=0)=
f^F(0,0)=\lim_{p^0\to \infty}f(0,0,p^0) .
\end{equation}
This means that the $\pi$ mesons with the momenta $k^3<O(p^0)$ and
$\vec{k}^{\bot}=0$ in the CM frame can be interpreted as the soft pion.
This fact holds even when the scaling violation effect exists,
since we can replace the exact scaling by the approximate one in this
discussion. In Weinberg's language, these soft pions correspond to
semi-soft pions \cite{wein}. The important point of this soft pion 
theorem is that the soft-pion limit can not be interchanged with the
manipulation to obtain the 
discontinuity of the reaction ``$a + b + \bar{\pi}_s \to a + b + \bar{\pi}_s$''. 
We must first take the soft pion limit in the reaction ``$a + b \to \pi_s + anything$''.
This is because the soft pion attached to the nucleon(anti-nucleon)
in the final state is missed in the discontinuity of the reaction
``$a + b + \bar{\pi}_s \to a + b + \bar{\pi}_s$'' where the soft pion limit is taken.\\
In Sect.2 we summarize detailed kinematics of the soft pion theorem in the
inclusive current-hadron reactions. In Sect.3 we review the previous works 
concerning the charge asymmetry and the contribution to the NMC deficit.
In Sect.4 we give the result for the combination of the
structure functions $(5F_2^{\nu N}/6 - 3F_{2}^{\mu N})$ which is discussed 
in the context of the charge symmetry violation of the parton distribution
functions\cite{adelaide} and can be regarded as the soft pion contribution 
to the strange sea quark in our case. In Sect.5 we give the soft-pion contribution
to the structure function $g_1$ and estimate its effect on the spin
dependent sum rules.  In Sect.6 conclusions are given.

\section{Kinematics}
\begin{figure}
   \centerline{\epsfbox{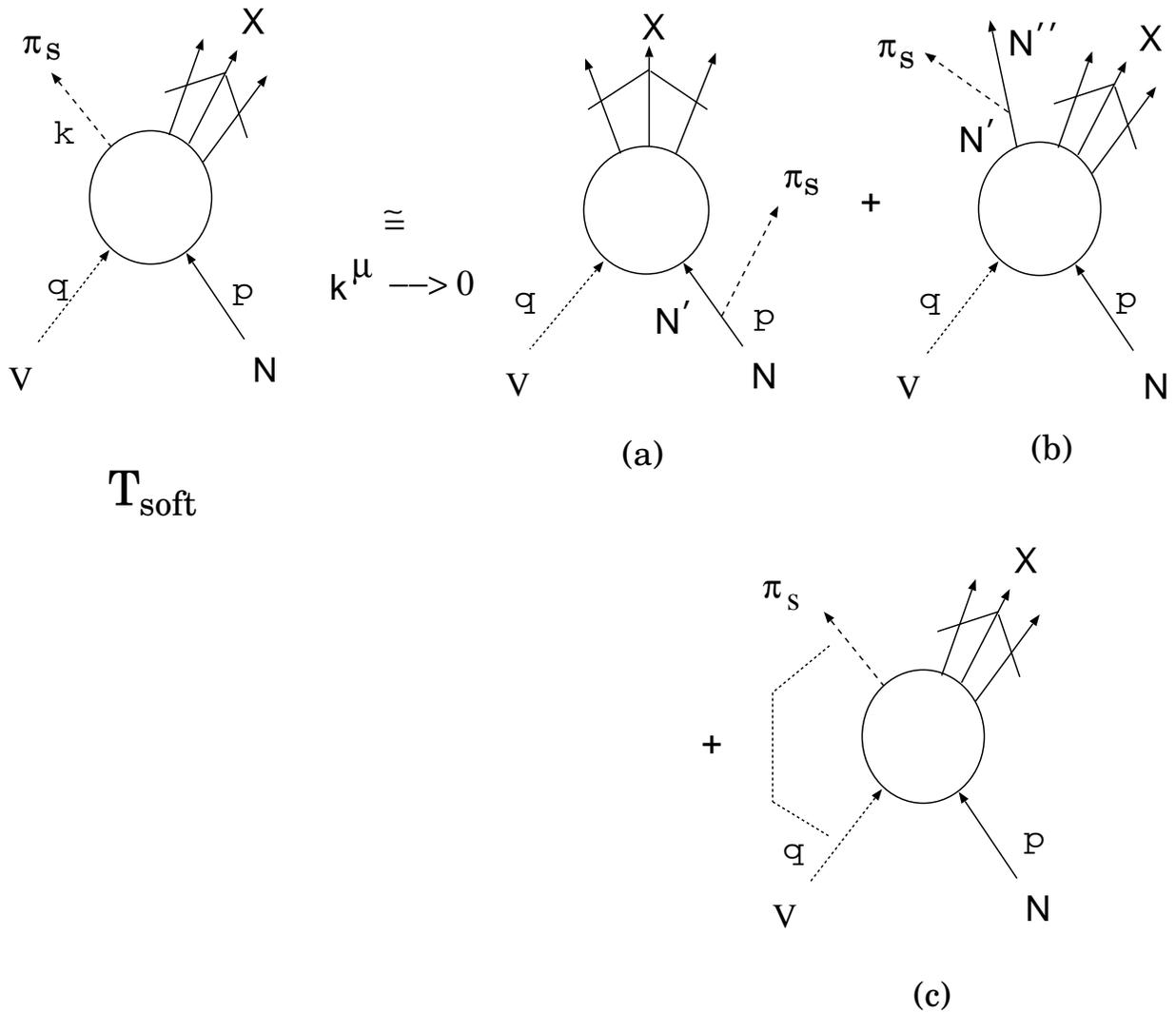}}
   \caption{Soft pion theorem in the inclusive reaction. the graph (a) corresponds
to the pion emission from the initial nucleon,the graph (b)corresponds to the pion
emission from the final nucleon (the anti-nucleon), and the graph (c) corresponds 
to the term coming from the null-plane commutator.}
        \label{fig:1}
\end{figure}
Let us consider the semi-inclusive current induced reactions 
$V_{a}^{\mu}(q) + N(p) \to \pi_{s}(k) + anythings(X)$\cite{kore78}.
The hadronic tensor of this reaction can be expressed as
\begin{equation}
T^{\mu \nu}=(m_{\pi}^2-k^2)^2\int d^4xd^4yd^4z\exp [-ik\cdot (x-z)+iq\cdot y]
\langle N(p)|[T^{\dagger}(\phi_{\pi}(x)V^{\mu}(y)),T(\phi_{\pi}(z)V^{\nu}(0))]
|N(p)\rangle_c ,
\end{equation}
where the spectral condition is used to express the tensor as the matrix element of 
the commutator. Then, we first take $k^{+}=0$ and $\vec{k}^{\bot}=0$, and after that
we take $k^{-}=0$. In this limit, $k^2$ is restricted to be 0, but the momentum of 
the initial particles are unrestricted. The amplitude in this limit is classified 
into three terms as in Fig.1. The graph (a) is the term where the proper part of 
the axial-vector current attached to the initial nucleon, the graph (b) is the
one to the final nucleon (anti-nucleon), and the graph (c) is the term
coming from the null-plane commutation relation at $x^{+}=0$. 
Using the PCAC relation
$\partial_{\mu}J_a^{5\mu}(x)=m_{\pi}^2F\phi_{\pi}(x)$, 
where $F=\sqrt{2}f_{\pi}$ for $a=1 \pm i2$ and $F=f_{\pi}$ for $a=3$,
the hadronic tensor in the soft pion limit can be classified as
\begin{equation}
W^{\mu \nu}_{abcd}=\sum_{i=1}^3 ((A_i^{\mu \nu}+B_i^{\mu \nu})
+C_1^{\mu \nu}+D^{\mu \nu}_4.
\end{equation}
Let us explain each term by taking $a=1+i2, c=1-i2,$ and $b=d^{\dagger}$. 
The term $A_1^{\mu \nu}$ corresponding to the graph (a) in Fig.2
is given as
\begin{eqnarray}
A^{\mu \nu}_1& =& \frac{1}{8f_{\pi}^2(p^+)^2}\int d^4y\exp (iq\cdot y)
\{\langle N(p)|J_a^{5+}(0)|N^{\prime}(p)\rangle 
\langle N^{\prime}(p)|V_b^{\mu}(y)V_d^{\nu}(0)|N^{\prime}(p)\rangle
\langle N^{\prime}(p)|J_c^{5+}(0)|N(p)\rangle \nonumber \\
&-&\{\langle N(p)|J_c^{5+}(0)|N^{\prime}(p)\rangle
\langle N^{\prime}(p)|V_d^{\nu}(0)V_b^{\mu}(y)|N^{\prime}(p)\rangle
\langle N^{\prime}(p)|J_a^{5+}(0)|N(p)\rangle \} .
\end{eqnarray}
The term $A_2^{\mu \nu}$ corresponding to the graph (b) in Fig.2 is given as
\begin{eqnarray}
A^{\mu \nu}_2& =& \frac{-1}{4f_{\pi}^2p^+}\int d^4x\int d^4y\exp (iq\cdot y)
\delta (x^+-y^+) 
\{\langle N(p)|[J_a^{5+}(x),V_b^{\mu}(y)]V_d^{\nu}(0)|N^{\prime}(p)\rangle 
\langle N^{\prime}(p)|J_c^{5+}(0)|N(p)\rangle \nonumber \\
&+& \langle N(p)|J_c^{5+}(0)|N^{\prime}(p)\rangle 
\langle N^{\prime}(p)|V_d^{\nu}(0)[J_a^{5+}(x),V_b^{\mu}(y)]|N(p)\rangle\} .
\end{eqnarray}
The terms $A_3^{\mu \nu}$ corresponding to the graph (c) in Fig.2 is given as
\begin{eqnarray}
A^{\mu \nu}_3& =& \frac{1}{4f_{\pi}^2p^+}\int d^4y\int d^4z\exp (iq\cdot y)
\delta (z^+)\{
\langle N(p)|J_a^{5+}(0)|N^{\prime}(p)\rangle
\langle N^{\prime}(p)|V_b^{\mu}(y)[J_c^{5+}(z),V_d^{\nu}(0)]|N(p)\rangle \nonumber \\
&+&\langle N(p)|[J_c^{5+}(z),V_d^{\nu}(0)]V_b^{\mu}(y)|N^{\prime}(p)\rangle 
\langle N^{\prime}(p)|J_a^{5+}(0)|N(p)\rangle \} .
\end{eqnarray} 
The term $B_1^{\mu \nu}$ corresponding to the graph (d) in Fig.2 is given as
\begin{eqnarray}
B_1^{\mu \nu}&=&\frac{1}{2f_{\pi}^2}\int d^4y\exp (iq\cdot
y)\sum_X\int \frac{d^3n}{(2\pi )^3(2n^+)^3} \nonumber \\
&\times &\{ \langle N(p)|V_b^{\mu}(y)|N^{\prime}(n)X\rangle 
\langle N^{\prime}(n)|J_a^{5+}(0)|N^{\prime \prime}(n)\rangle 
\langle N^{\prime\prime}(n)|J_c^{5+}(0)|N^{\prime}(n)\rangle
\langle N^{\prime}(n)X|V_d^{\mu}(0)|N(p)\rangle \nonumber \\
&- & \langle N(p)|V_d^{\nu}(y)|N^{\prime}(n)X\rangle 
\langle N^{\prime}(n)|J_c^{5+}(0)|N^{\prime \prime}(n)\rangle 
\langle N^{\prime \prime}(n)|J_a^{5+}(0)|N^{\prime}(n)\rangle
\langle N^{\prime}(n)X|V_b^{\nu}(0)|N(p)\rangle \} ,
\end{eqnarray}
where $N^{\prime}$ and $N^{\prime \prime}$ mean the sum over the possible nucleon or the anti-nucleon.
The term $D^{\mu \nu}$ corresponding to the graph (i) in Fig.2 is given by
\begin{eqnarray}
D^{\mu \nu}_4&=&-\frac{1}{2f_{\pi}^2}\int d^4xd^4yd^4z\exp (iq\cdot y)
\delta (x^+ - y^+)\delta (z^+) \langle
N(p)|[[J_a^{5+}(x),V_b^{\mu}(y)],[J_c^{5+}(z),V_d^{\nu}(0)]]|N(p)\rangle
\end{eqnarray}
The terms $B_2^{\mu \nu}, B_3^{\mu \nu}, C_1^{\mu \nu}$
corresponding to the graphs (e),(f) and ((g)+(h)) respectively
are discarded in the deep-inelastic region by the following reason:
These graphs are characterized by the one soft pion emission from the final 
nucleon (anti-nucleon), and this has the matrix element of the following form
\begin{equation}
\langle p(p),h|J^{5+}_{a}(0)|n(p),h^{\prime}\rangle = 2p^+hg_A(0)\delta_{hh^{\prime}},
\end{equation}
where $p$ means the proton and $n$ means the neutron and we take
$a=1 + i2$ by way of illustration. We have the helicity factor $h$
in the matrix element.
At high energy, because of this factor we can expect that the contribution from 
the (+) helicity nucleon (anti-nucleon)
and that of the (-) helicity one cancels each other.
While, at low energy, these graphs are suppressed by the
form factor effect in the deep-inelastic region. Hence contributions
from these graphs are expected to be small compared with those from other graphs.\\
Now we define the $W^{\mu \nu}_{\alpha \beta}$ as
\begin{equation}
W^{\mu \nu}_{\alpha \beta}=\frac{1}{4\pi}W^{\mu \nu}_{abcd},
\end{equation}
\begin{figure}
   \centerline{\epsfbox{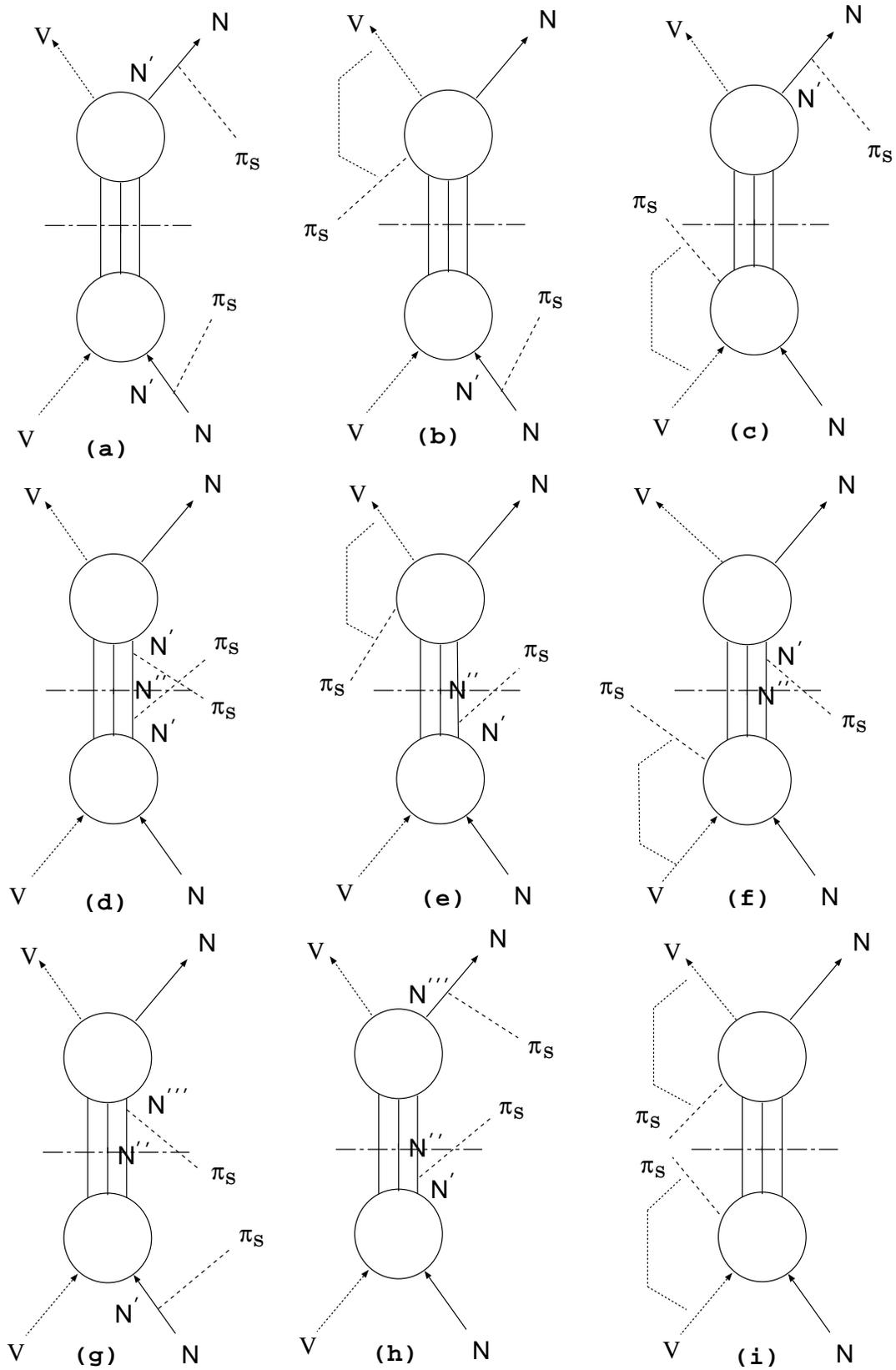}}
   \caption{The graph contributing to the hadronic tensor}
        \label{fig:2}
\end{figure}
and the parity-conserving parts of the structure function of the semi-inclusive soft pion reaction as
\begin{equation}
W^{\mu \nu}_{\alpha \beta}=(-g^{\mu \nu}+\frac{q^{\mu}q^{\nu}}{q^2})W_{1\gamma}^{\alpha \beta} + 
(p^{\mu}-\frac{\nu}{q^2}q^{\mu})(p^{\nu}-\frac{\nu}{q^2}q^{\nu})W_{2\gamma}^{\alpha \beta}
+i\epsilon_{\mu \nu \lambda \sigma}q^{\lambda}s^{\sigma}G_{1\gamma}^{\alpha \beta}
+i\epsilon_{\mu \nu \lambda \sigma}q^{\lambda}(s^{\sigma}\nu -q\cdot sp^{\sigma})G_{2\gamma}^{\alpha \beta},
\end{equation}
where $\nu = p\cdot q$.
We define the structure functions $F_{1\gamma}^{\alpha \beta}, F_{2\gamma}^{\alpha \beta}, 
g_{1\gamma}^{\alpha \beta}, g_{2\gamma}^{\alpha \beta}$ as
$W_{1\gamma}^{\alpha \beta}=F_{1\gamma}^{\alpha \beta},
\nu W_{2\gamma}^{\alpha \beta}=F_{2\gamma}^{\alpha \beta},
\nu G_{1\gamma}^{\alpha \beta}=g_{1\gamma}^{\alpha \beta}, 
\nu^2G_{2\gamma}^{\alpha \beta}=g_{2\gamma}^{\alpha \beta}$.
Since the soft pion momentum is zero, the structure of the hadronic tensor is the same as
in the total inclusive reaction. Here the suffix $\alpha$ specifies the charge of the
boson coupled to the current, the suffix $\beta$ specifies the pion charge, and
the suffix $\gamma$ specifies the target nucleon.
In case of the virtual photon, we discard the suffix $\alpha$. Thus, in this case, we denote
the hadronic tensor as $W^{\mu \nu}_{\beta}$ and the corresponding structure 
functions as $F_{i\gamma}^{\beta} (g_{i\gamma}^{\beta})$ for $i=1,2$.

\section{The charge asymmetry and the contribution to the Gottfried sum from the soft pion}
The contribution from the terms $A_1^{\mu \nu}$ and $B_1^{\mu \nu}$ corresponding to
the graphs (a) and (d) in Fig.2 can be related to the known process directly. 
However the contributions from the terms $A_2^{\mu \nu}, A_3^{\mu \nu}$ and $D_4 ^{\mu \nu}$
are not directly related to the known process. We need some methods to calculate
these parts. Since we consider the deep inelastic limit, all contributions are light-cone
dominated as is easily checked by the explicit expression in the previous section.
Hence we can use here the cut vertex formalism\cite{Mue}. 
Then to obtain the structure function we need to invert the moment sum rules.
Here a tacit assumption is introduced and the result is simply the relation to
the structure function in the total inclusive reactions. What we do in such a
discussion is how the $Q^2$ dependence enters, and the relation between the
structure functions are unchanged by this $Q^2$ dependence. Because of this fact,
we use the light-cone current algebra\cite{fg} at some initial $Q^2=Q_0^2$ where
the evolution is started, and use the symmetry relation embodied in this algebra
by taking care of the fact whether it is a singlet piece or a non-singlet piece
and whether it is a charge conjugation even piece or a odd piece.  The latter point is
expressed as the symmetric bilocal or the antisymmetric bilocal in the light-cone
current algebra. These classification is necessary because they have different
$Q^2$ dependence. Then, once we have the relation to the structure function in the
total inclusive reaction at $Q_0^2$, the $Q^2$ dependence enters following the
$Q^2$ dependence of the structure function related by this way.\\
Now the method in Ref.\cite{sakai} had not been checked experimentally, hence it was
done in the soft $\pi^{-}$ case\cite{kore78reso} in the electroproduction. 
From the experimental data of the Harvard-Cornell group\cite{harvard} the data 
satisfying the following conditions are selected.
\begin{enumerate}
\item[(1)]The transverse momentum satisfies $|\vec{k}^{\bot}|^2\leq m_{\pi}^2$.
\item[(2)]The change of $F^-$ can be regarded to be small in the small $x_F$ region.
\end{enumerate}
The effective cut of $x_F$ under these two conditions 
is about $0.2$. Then the theoretical value 
is about 10\% $\sim$ 20\% of the experimental value. However,
in the central region, there are many pions from the decay of the resonances, and about
20\% $\sim$ 30\% can be expected to be the pion from the directly produced pion.
Hence the theoretical value is the same order with the experimental value.
Now to reduce the ambiguity due to the pion from the resonance decay product,
the charge asymmetry was calculated\cite{kore82}. This is because the pion from
the resonance decay is charge symmetric as far as the resonance production and the 
decay succeeding it are governed by the strong interaction. In fact this is the reason
why we consider the particle in the central region is charge symmetric.
Thus if charge asymmetry exists in such a kinematical region, there should be some clear 
physics behind it. The soft pion theorem is one candidate for this, since the pole term 
is asymmetric in charge. For example, in the proton target case, the terms contributing to 
the soft $\pi^-$ production are given by the graphs in Fig.3, while those of 
the soft $\pi^+$ are given by the graphs in Fig.4. 
\begin{figure}
   \centerline{\epsfbox{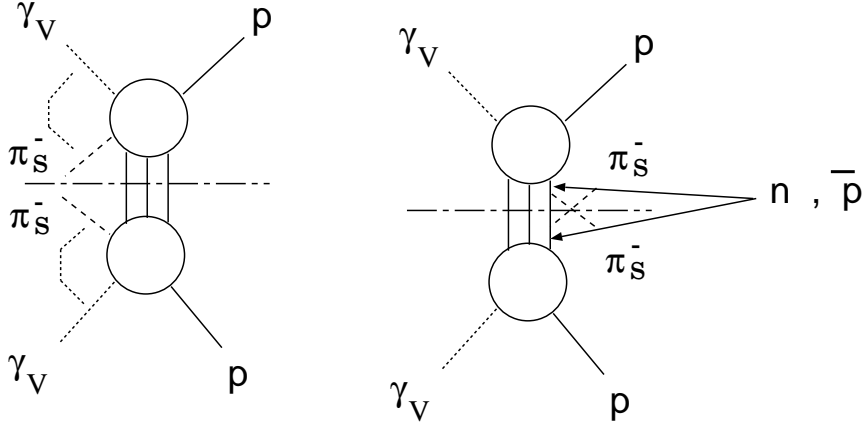}}
   \caption{The graphs contributing to the soft $\pi^-$ production in the proton target case.}
        \label{fig:3}
\end{figure}
\begin{figure}
   \centerline{\epsfbox{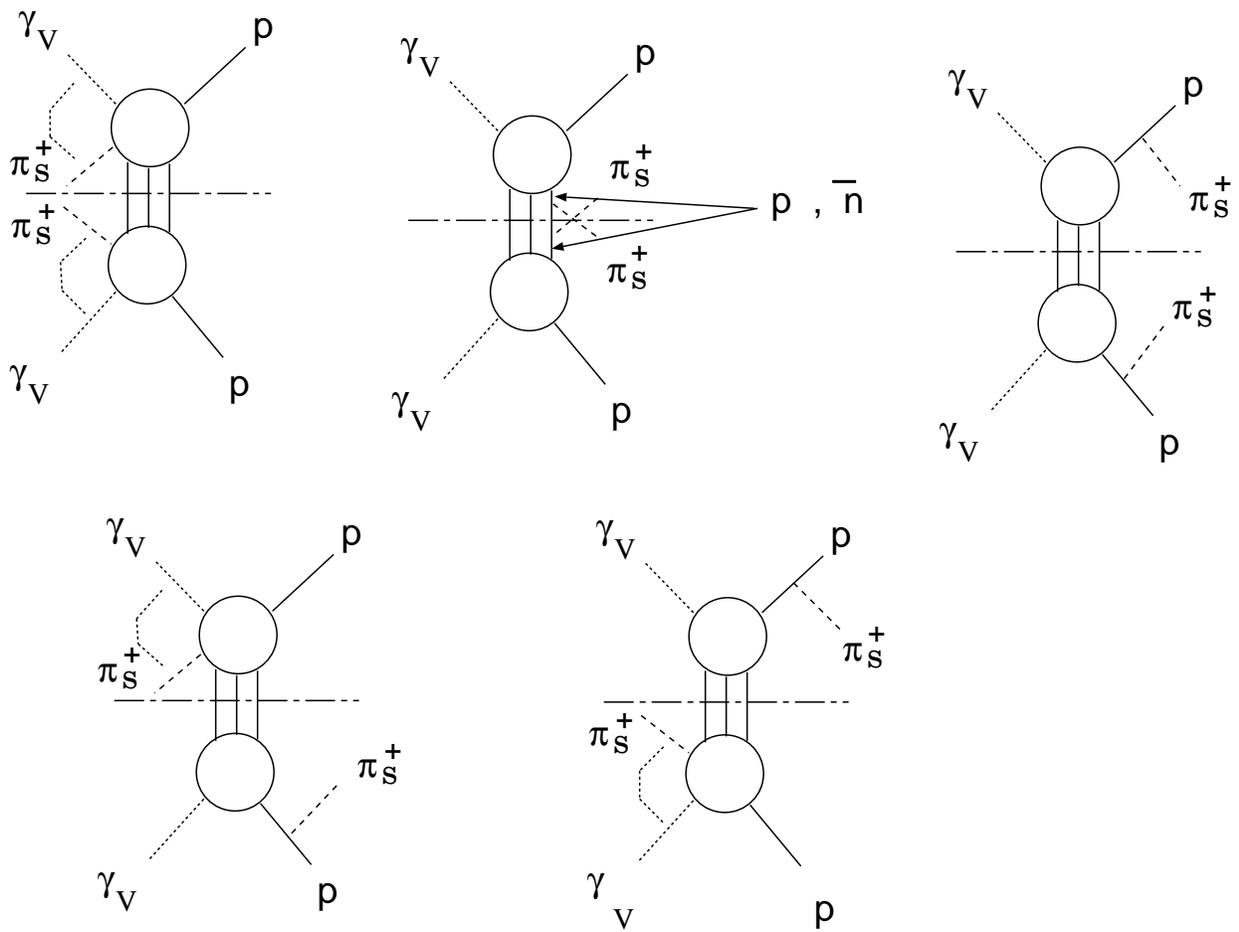}}
   \caption{The graphs contributing to the soft $\pi^+$ production in the proton target case.}
        \label{fig:4}
\end{figure}
By assuming symmetric sea polarization for simplicity, we obtain\cite{kore82}
\begin{eqnarray}
F_{2p}^+ - F_{2p}^- & =\frac{1}{4f_{\pi}^2}[F_2^{\bar{\nu}p}-F_2^{\nu p} +2g_A^2(0)F_2^{en}
-8xg_A(0)(g_1^{ep}-g_1^{en})\nonumber \\
& + 2g_A^2(0)(<n>_p + <n>_{\bar{n}}-<n>_{\bar{p}}-<n>_n)F_2^{ep}],
\end{eqnarray}
\begin{eqnarray}
F^+ - F^- & = \frac{1}{64\pi^3f_{\pi}^2}[\frac{F_2^{\bar{\nu}p}-F_2^{\nu p}}{F_2^{ep}}
+2g_A^2(0)\frac{F_2^{en}}{F_2^{ep}}-8xg_A(0)\frac{(g_1^{ep}-g_1^{en})}{F_2^{ep}} \nonumber \\
& + 2g_A^2(0)(<n>_p + <n>_{\bar{n}}-<n>_{\bar{p}}-<n>_n)],
\end{eqnarray}
where $F^{\alpha}$ is defined as
\begin{equation}
F^{\alpha}=\frac{1}{\sigma_T}q^0\frac{d\sigma^{\alpha}}{d^3q},
\end{equation}
and the suffix in the structure function of the total inclusive reaction is given  
to show the reaction concerned. Without the symmetric sea polarization assumption,
$(g_1^{ep}-g_1^{en})$ in Eq.(14) should be changed to $\displaystyle{\frac{3}{2}(g_1^{ep}-g_1^{en})
-\frac{1}{12}(g_1^{\bar{\nu}p}-g_1^{\nu n})}$.
By neglecting the nucleon multiplicity term which can be expected to be a small positive
contribution, the theoretical value is roughly equal to $0.15 \sim 0.18$ with a week $x_B$
dependence. While the experimental value with the transverse momentum satisfying
$|\vec{k}^{\bot}|^2\leq m_{\pi}^2$ in Ref.\cite{harvard} 
is almost constant in the region $ 0< x_F < 0.1$ with its value $0.28 \pm 0.05$,
and it gradually decreases above $x_F = 0.1$. The data also shows a week $x_B$ dependence.
Hence the theoretical value is very near to the experimental value.\\
Based on these investigations, the contribution to the Gottfried sum was estimated\cite{kore99}. 
Adding the contributions from the soft $\pi_s^+,
\pi_s^-,$ and $\pi_s^0$, and subtracting the contributions to
$F_2^{en}$ from those to $F_2^{ep}$, we obtain
\begin{eqnarray}
\lefteqn{(F_2^{ep} - F_2^{en})|_{soft}}\nonumber \\
 &=& \frac{I_{\pi}}{4f_{\pi}^2}[g_A^2(0)(F_2^{ep} - F_2^{en})(3<n> -1)
-16xg_A(0)(g_1^{ep} - g_1^{en})] ,
\end{eqnarray}
where $I_{\pi}$ is the phase space factor for the soft pion defined as
\begin{equation}
I_{\pi} = \int\frac{d^2\vec{k}^{\bot}dk^+}{(2\pi)^{3}2k^+} 
\end{equation}
where $<n>$ is the sum of the nucleon and anti-nucleon multiplicity defined as
$<n>=<n>_p + <n>_n + <n>_{\bar{p}} + <n>_{\bar{n}}$.
In Eq.(5), the contribution coming from $D^{\mu \nu}_4$ cancels
out, among the terms proportional to $g_A^2(0)$ the one which has 
a factor $<n>$ comes from $B_1^{\mu \nu}$ and the other one
comes from $A_1^{\mu \nu}$, and the term proportional to
the spin dependent function $(g_1^{ep} - g_1^{en})$ comes from
$(A_2^{\mu \nu}+A_3^{\mu \nu})$. Note that this spin dependent term
is obtained in the symmetric sea polarization approximation. 
Without this approximation, $(g_1^{ep} - g_1^{en})$ in Eq.(17) should be replaced by
$\displaystyle{\frac{3}{2}(g_1^{ep} - g_1^{en})-\frac{1}{12}(g_1^{\bar{\nu} p} - g_1^{\nu p})}$
as in the charge asymmetry case. \\
Now the magnitude of the soft pion contribution depends largely on the phase 
space factor $I_{\pi}$ given by Eq.(18). To estimate this, information
of the experimental check of the charge asymmetry discussed first in this
section is used. We regard the directly produced pions in the virtual-photon 
and the target nucleon CM frame which satisfies the two conditions as the soft pion.
\begin{enumerate}
\item[(1)]The transverse momentum satisfies $|\vec{k}^{\bot}|\le bm_{\pi}$.
\item[(2)]The Feynman scaling variable $x_{F}=2k^3/\sqrt{s}$ satisfies $|x_{F}|\le c$.
\end{enumerate}
Here $s$ is defined as $s=(p+q)^2$, and we take the value of $b$ near 1, and that
of $c$ near 0.1 because the charge asymmetry is explained by these values
fairly well. The effect of the change of these parameters are discussed 
in\cite{kore99}. We parameterize the nucleon multiplicity as
\begin{equation}
\langle n \rangle = a\log_es +1  ,
\end{equation}
where a is fixed to be $0.2$ in consideration of the proton ant the
anti-proton multiplicity in the $e^+e^-$ annihilation such that
$a\log_e\sqrt{s}$ with $\sqrt{s}$ replaced by the CM energy of that
reaction agrees with the multiplicity of that reaction\cite{DELPHI}.
Further, for an explicit evaluation of the $(F_2^{ep} - F_2^{en})$
and $(g_1^{ep} - g_1^{en})$ on the right-hand side of Eq.(17),
we approximate them by the symmetric sea quark distributions both
for the unpolarized and the polarized ones and use the parameter given
by MRS and GS\cite{GS} at $Q_0^2=4$GeV$^2$. An effect of the inclusion
of the asymmetry is small in the $x$ region considered compared with the 
ambiguity of the factor $I_{\pi}$. After all 
the magnitude of the soft pion contribution to
the Gottfried sum is found to be $-0.04 - -0.02$.\\
Undoubtedly, the soft pion contributes to the structure function.
The crucial point is whether it can be neglected or not. The discussion here
shows that it cannot be neglected. In the parton model, 
the soft pion contribution is overlooked, because it violates the
generalized unitarity\cite{sakai} and no particular attention is paid
on this point. We must be very careful to take the soft pion limit
in the unitarity relation. In other words, we have not yet
succeeded to express the intermediate state of the unitarity sum in the
hadronic reaction only by the quarks and gluons. Hence the
non-perturbative effects such as the spectral condition of the hadron
,soft pion effects, and so on should be taken into account effectively.
In this point, soft pion theorem in the inclusive reaction is very 
particular by the aforementioned reason. Now, physical observables
are structure functions, and hence, if we try to express the
experimental values by the quark distribution functions,
the soft pion contribution is effectively incorporated to these 
distributions. The Adler sum rule is a very general constraint,
and the valence quark distribution functions are restricted by
this sum rule severely. The soft pion contributes to this sum rule,
hence the valence quark distribution functions which satisfy
$\int_0^1dx\{u_v(x) - d_v(x)\}=1$ effectively take into account
this contribution. In the Gottfried sum, the valence quarks enter
with this combination, hence the soft pion contribution to the
Gottfried sum should be taken into the sea quark distributions.
Thus Eq.(17) can be expressed by the soft pion contribution to the
sea quark as
\begin{eqnarray}
\lefteqn{x(\lambda_{\bar{d}} - \lambda_{\bar{u}})|_{soft}}\nonumber \\
 &=& - \frac{I_{\pi}}{8f_{\pi}^2}[g_A^2(0)(F_2^{ep} - F_2^{en})(3<n> -1)
-16xg_A(0)(g_1^{ep} - g_1^{en})] ,
\end{eqnarray}
where $\lambda_{i}(\lambda_{\bar{i}})$ is a sea quark distribution function 
of the $i$ th quark with $i=u,d,s$, and we assume $\lambda_{i}=\lambda_{\bar{i}}$.
We give a plot of $x(\lambda_{\bar{d}} - \lambda_{\bar{u}})$
in the case of $a=0.2,b=1,c=0.1$ in Fig.5 along with the CTEQ4M fit of this
distribution\cite{CTEQ}.  
\begin{figure}
   \centerline{\epsfbox{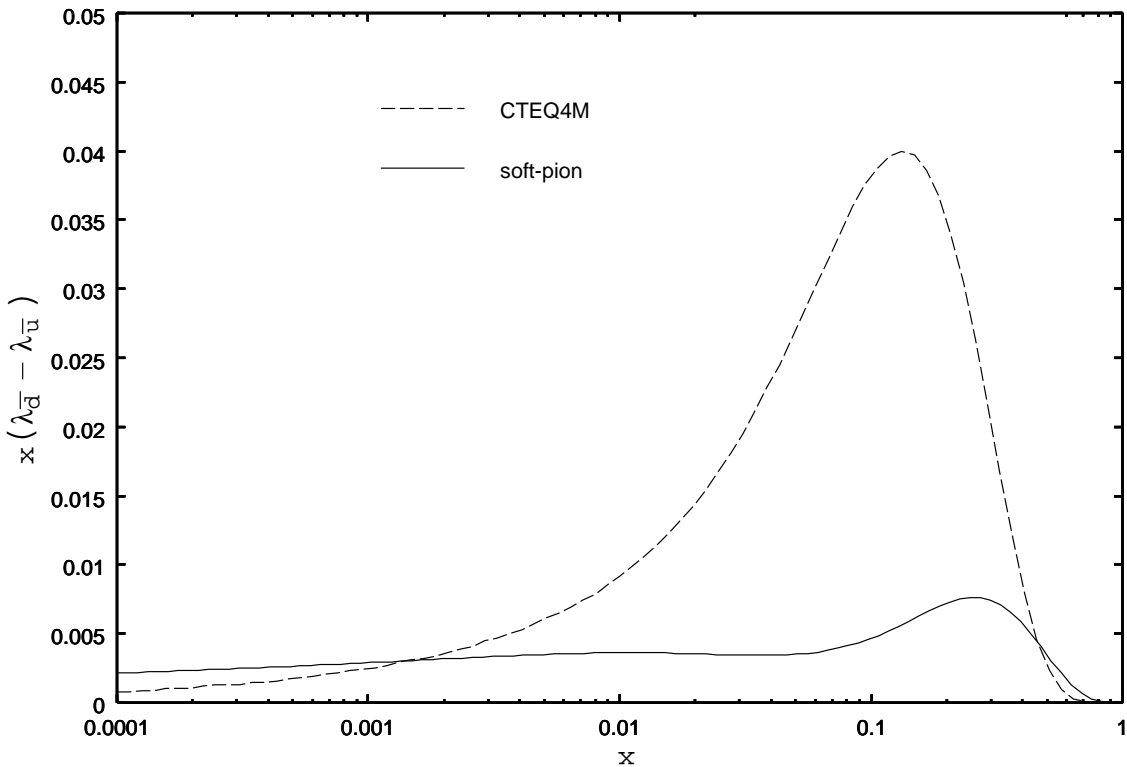}}
   \caption{The soft pion contribution to the sea quarks}
        \label{fig:5}
\end{figure}
The contribution of the soft pion to the Gottfried sum by these parameters
are about $-0.03$. As explained in the introduction, 
the contribution of this amount is
just the one required by the typical calculation by the meson
cloud model\cite{meso}. Further the fact that the main contribution
of this piece comes from the high energy region is consistent
with the analysis based on the modified Gottfried sum rule\cite{Got}.
Since the modified Gottfried sum rule holds in a very general context,
if we parameterize the sea quark distributions to satisfy the sum rule,
we can effectively take into account the non-perturbative effects
such as the soft pion. The situation is similar to the valence
quark distributions which satisfy the Adler sum rule. 
Now we have many sum rules which
have a clear physical meaning\cite{kore98,kore-ichep}. All these
sum rules give us constraints on the sea quarks. Unfortunately,
these sum rule constraints are disregarded in almost all the presently available
quark distribution functions except the constraint from the
Gottfried sum. This point is discussed in Ref.\cite{kore98} and is
shown that no strange sea quark distribution exists which satisfies the
sum rule constraint. In a recent analysis of the charge symmetry
violation, it is found that the symmetry violation is reduced
greatly and that the residual symmetry violation depends
on the resolution of the uncertainty on our knowledge of the strange sea quark
distribution\cite{csymv}.  Thus we study the point by calculating the
soft pion contribution to the strange sea quarks
 
\section{soft pion contribution to the strange sea quark}
Let us first consider the mean charge sum rule for the sea quarks in the proton
given in ref.\cite{kore98,kore-ichep} as
\begin{equation}
\frac{1}{3}\int_0^1dx\{ 2\lambda_u - \lambda_d - \lambda_s \} = 0.23.
\end{equation}
If we use the usual ansatz for the strange sea quark
\begin{equation}
\lambda_s = \frac{1}{4}(\lambda_u + \lambda_d),
\end{equation}
together with the assumption that the up sea quark and the down sea quark 
in the very small $x$ region is almost symmetric,
we can easily understand that the left-hand side of Eq.(21) diverges, and that it contradicts 
with the right-hand side of it. The fact that the up sea quark and the down sea quark 
in the very small $x$ region becomes almost symmetric is one of the results in the sum rule
approach which stems from the assumption where the pomeron is flavor singlet. Moreover, 
from the sum rule approach, we have the result that the sea quarks are flavor singlet in the 
very small $x$ region under the same assumption. 
Thus the ansatz (22) is wrong. We know that this relation is satisfied
except in the very small $x$ region from the phenomenological analysis. 
Hence there is a region where this relation breaks down. To see how far we can use the ansatz (22),
we use the usual parameterization given by \cite{CTEQ}, and estimate the sum
rule (21), and find that the ansatz (22) should be abandoned at least below $x=1\times 10^{-3}$,
since the contribution below this region becomes very large. Thus, below the 
region $x=1\times 10^{-1}$, there is a transition
region from the ansatz (22) to the symmetric distribution 
though it is not clear whether it is gradually or sharply. What is clear is
that we would encounter something strange in the small $x$ region below $x=1\times 10^{-1}$
as far as we keep the ansatz (22). In this sense it is interesting that the phenomenological
indication of the contradiction is reported in the small $x$ region\cite{adelaide},
though its magnitude of the contradiction is reduced greatly\cite{csymv}. 
We consider that the phenomenon may be related to the soft pion production at high energy 
which gives seeming charge symmetry violation effects.
We define
\begin{equation}
F_2^{\mu N}= \frac{1}{2}(F_2^{\mu p} + F_2^{\mu n}) ,\hspace{1cm} 
F_2^{\nu N}= \frac{1}{4}(F_2^{\nu p} + F_2^{\nu n} + F_2^{\bar{\nu} p} + F_2^{\bar{\nu} n} ) .
\end{equation}
Then the soft pion contribution to the following combination of the
structure functions with the Cabibbo angle being 0 is given as
\begin{eqnarray}
(\frac{5}{6}F_2^{\nu N} - 3F_2^{\mu N})|_{soft} = \frac{I_{\pi}}{f_{\pi}^2}[\frac{1}{12}(F_2^{\nu p}
+ F_2^{\bar{\nu} p})_0 + \frac{5}{16}g_A^2(0)(F_2^{\nu p}+ F_2^{\bar{\nu} p})(1+\langle n \rangle^{\nu N})\\\nonumber
-\frac{9}{8}(F_2^{ep} + F_2^{en})g_A^2(0)(1+\langle n \rangle^{eN}) - 10xg_A(0)(g_1^{ep} - g_1^{en})],
\end{eqnarray}
where $(F_2^{\nu p}+ F_2^{\bar{\nu} p})_0$ is the structure function in the $SU(3)$.
By neglecting the heavy quark contribution
such as the charm, the left hand side of Eq.(24) is proportional to the strange sea quark
distribution. Hence, under the same approximation, we express the right hand side of Eq.(24) 
only by the light sea quark distribution functions, and obtain
\begin{eqnarray}
x\lambda_s|_{soft} = \frac{I_{\pi}}{f_{\pi}^2}[\frac{x}{6}(d_v + u_v)
+ \frac{x}{3}(\lambda_u + \lambda_d) + \frac{3}{4}xg_A^2(0)(1+\langle n \rangle)\lambda_s \\\nonumber
- \frac{5}{3}xg_A(0)(\triangle u_v - \triangle d_v)],
\end{eqnarray} 
where we set $\langle n \rangle^{\nu N}=\langle n \rangle^{eN}=\langle n \rangle$.
By using the same parameters as in the previous section,i.e.,the parameters $a=0.2,b=1,c=0.1$ and
the distribution by MRS and GS\cite{GS} at $Q_0^2=4$GeV$^2$,  we plot the soft pion contribution
to the strange sea quark in Fig.6.
\begin{figure}
   \centerline{\epsfbox{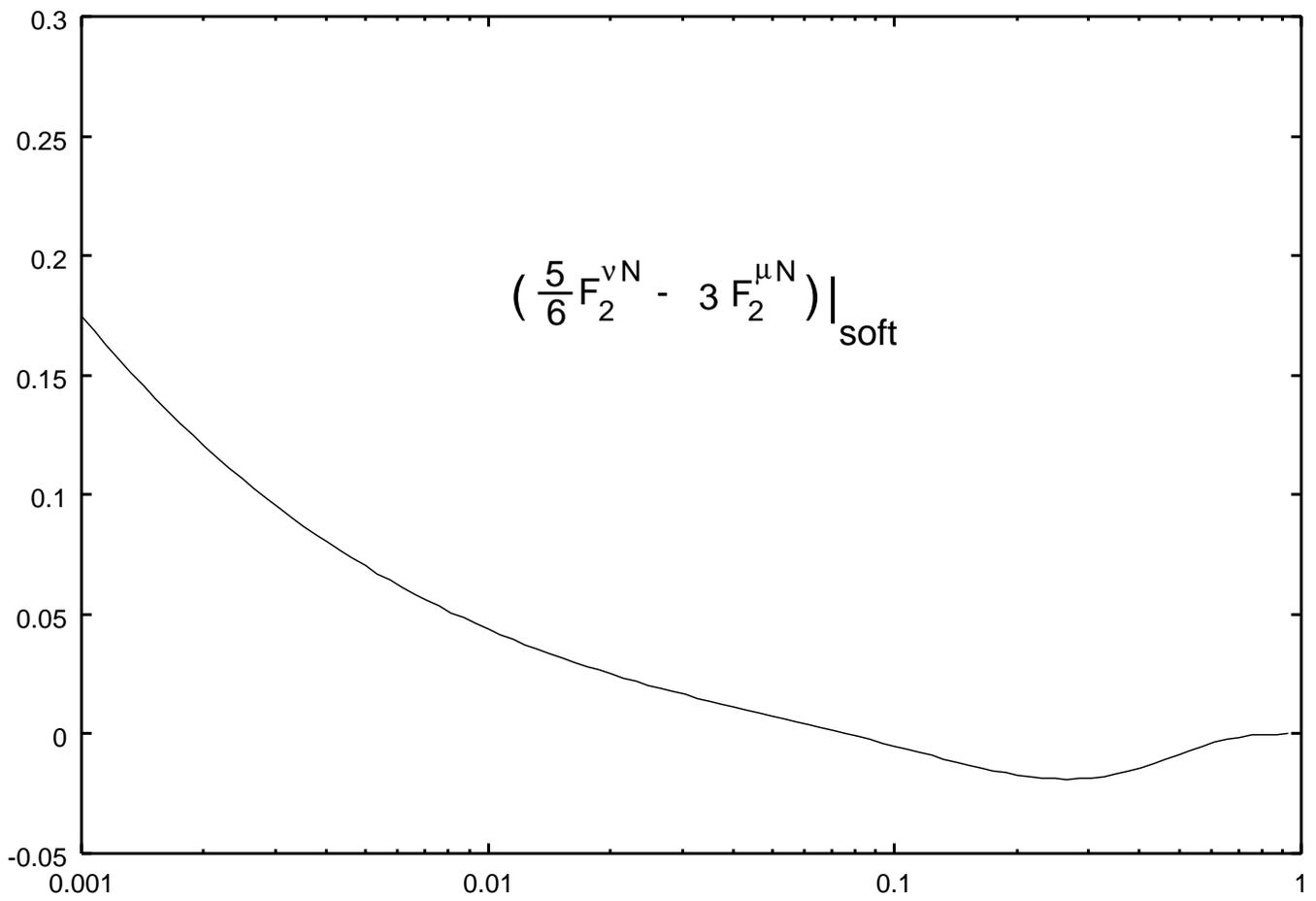}}
   \caption{The soft pion contribution to the $x\lambda_s|_{soft}=(\frac{5}{6}F_2^{\nu N} - 3F_2^{\mu N})|_{soft}$
in the proton target.}
        \label{fig:6}
\end{figure}
From Fig.6, we see that the soft pion contributes below $x\sim 0.1$ and its magnitude is 
similar to the one discussed in \cite{csymv}. Further we see that the strange sea quark 
where the soft pion contribution is added to the ansatz (22) greatly improves the convergence 
of the sum rule (21). In fact the numerical integration of the left hand side of 
the sum rule (21) with use of the ansatz (22) is 0.26 from $x=1\times 10^{-3}$ to 1
and 0.60 from $x=1\times 10^{-4}$ to 1. It goes without saying that the contribution
below $1\times 10^{-4}$ is very large and it ultimately diverges in this case.
On the other hands, if we add the soft pion contribution to the strange sea
quark distribution given by the ansatz (22), this numerical integration becomes 0.18 
from $x=1\times 10^{-3}$ to 1 and 0.27 from $x=1\times 10^{-4}$ to 1. 
This is because the modified strange sea quark distribution rapidly reaches the symmetric one
below $x=1\times 10^{-3}$ just as the sum rule approach predicts. Thus the sum rule can be 
expected to be well satisfied by this modification if we extrapolate the distributions 
in the very small $x$ region to the symmetric one once the modified strange sea quark
reaches it.

\section{soft pion contribution to the spin dependent structure function $g_1$}
The soft pion contribution to the structure function
$g_1^{ep}$ and $g_1^{en}$ are given as
\begin{eqnarray}
xg_1^{ep}|_{soft}
=&\frac{I_{\pi}}{4f_{\pi}^2}[-x(g_1^{\nu p}+g_1^{\bar{\nu} p})+g_A^2(0)x(g_1^{ep}+2g_1^{en})
+3g_A^2(0)<n>xg_1^{ep} \nonumber \\
&-3g_A(0)(F_2^{ep} - F_2^{en})-\frac{g_A(0)}{6}(F_2^{\nu p} - F_2^{\bar{\nu} p})],
\end{eqnarray}
and 
\begin{eqnarray}
x(g_1^{ep}-g_1^{en})|_{soft}
=&\frac{I_{\pi}}{4f_{\pi}^2}[g_A^2(0)(3\langle n \rangle -1)x(g_1^{ep}-g_1^{en})
-6g_A(0)(F_2^{ep} - F_2^{en})-\frac{g_A(0)}{3}(F_2^{\nu p} - F_2^{\bar{\nu} p})].
\end{eqnarray}
By expressing the right hand side of these equations by the quark distribution
functions with the symmetric sea distributions both for the polarized and the
unpolarized ones and using the same parameters as in the previous sections, 
it is straightforward to estimate the contribution from the soft pion to the 
Bjorken sum rule and the Ellis-Jaffe sum rule. 
We find that the contribution from the region above $x=1\times 10^{-4}$
are negligible small, however we find that $g_1^{ep}$ itself becomes large 
below $x=1\times 10^{-3}$. Hence we give a plot of it in Fig.7, and the
ratio of the soft pion contribution to the input Gehrmann and Stirling
distribution at $Q^2=4$GeV$^2$ in Fig.8.
\begin{figure}
   \centerline{\epsfbox{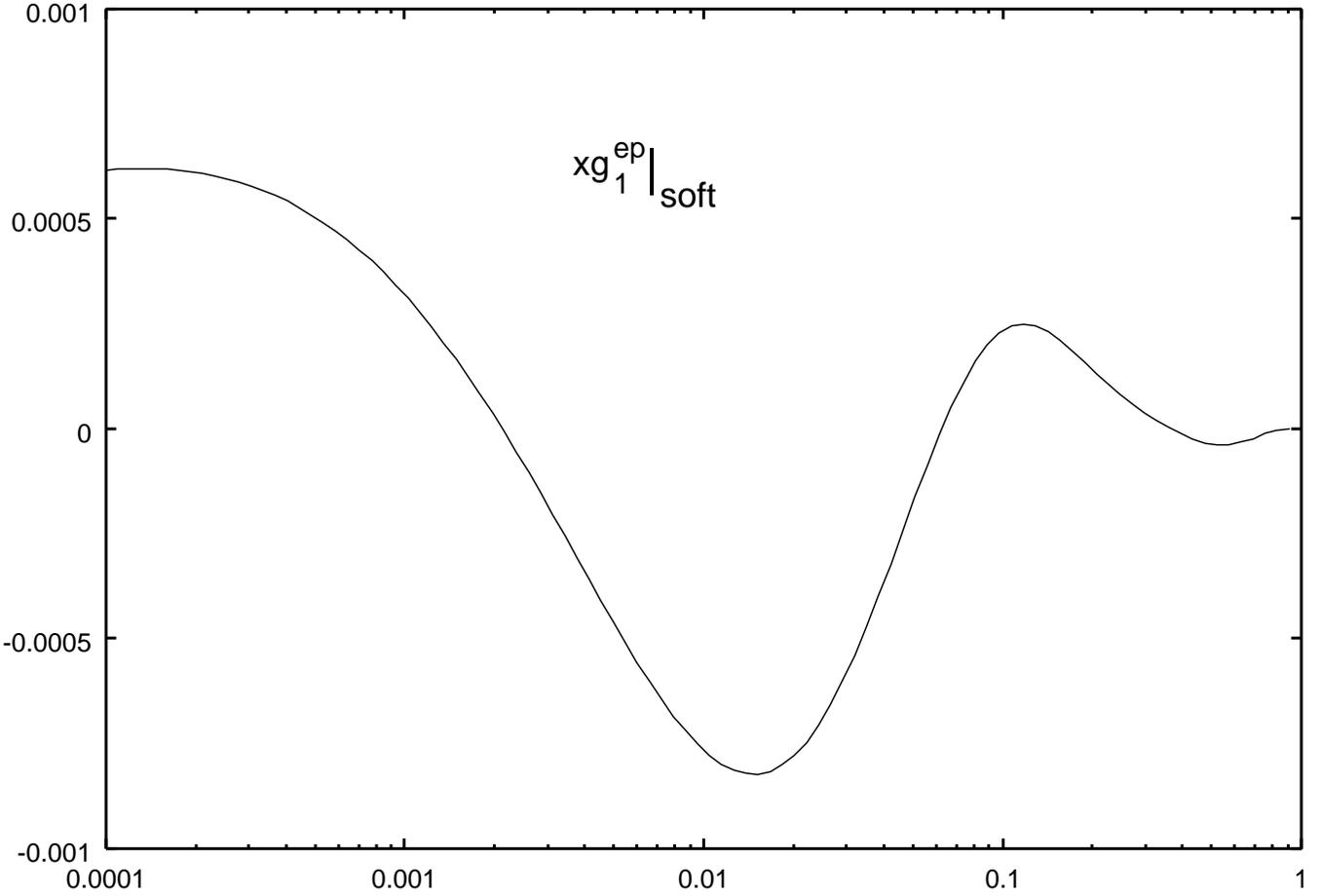}}
   \caption{The soft pion contribution to the $xg_1^{ep}$.}
        \label{fig:7}
\end{figure}
\begin{figure}
   \centerline{\epsfbox{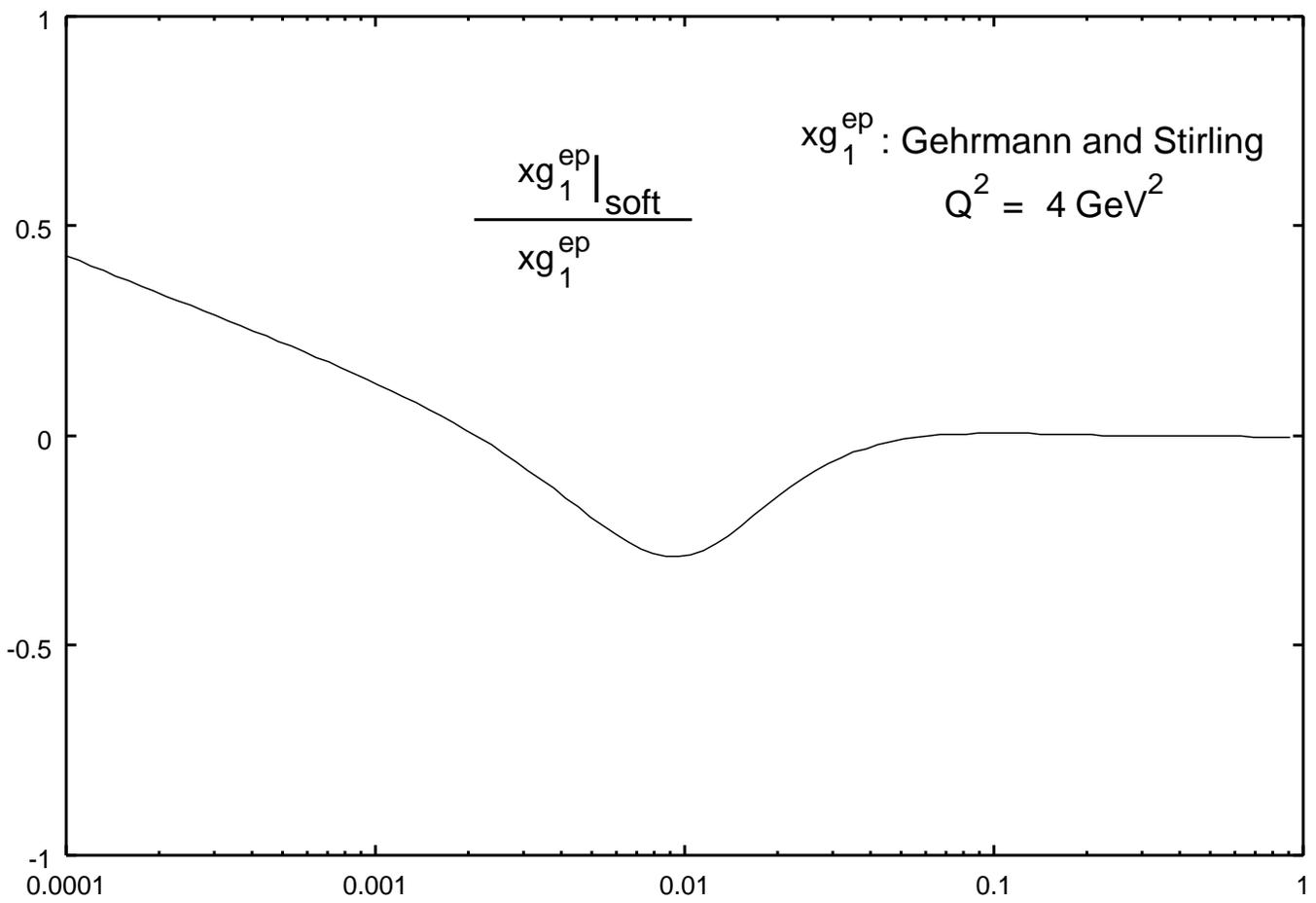}}
   \caption{The ratio of the soft pion contribution 
to the Gehrmann and Stirling distribution at $Q^2=4$GeV$^2$.}
        \label{fig:8}
\end{figure}
\section{Conclusion}
When the soft pion contributes to the structure functions, we must be careful
to take it as far as its magnitude being non-negligible. At high energy,
this component is overlooked in almost all models. However, if the Feynman scaling 
is satisfied, this piece can not be neglected as explained in this paper. 
Phenomenologically, we observe its effect as if the charge symmetry is violated. 
This originates from the fact that the pole term of the soft pion contribution is 
asymmetric in charge. Hence this is not the symmetry violation but the remnant of the
spontaneous chiral symmetry breakings. In some cases,
we have taken this piece unconsciously. This happens, for example, in the case
when we make the quark distribution to satisfy the sum rule
as in the the case of the Adler sum rule. In this sense, the sum
rule constraint is very important. Now, the Gottfried sum is the experimentally determined 
quantity, and phenomenologically determined sea quarks to satisfy this sum 
include the soft pion contribution.
The original Gottfried sum rule\cite{Gottfried} has failed to give the correction to $1/3$, 
since the quantity ($\infty - \infty $) has not been considered properly,i.e., simply
disregarded.  The method based on the current anticommutation relation on the null-plane 
gives us the way to separate the part which gives $\infty$, hence we can fix
the quantity ($\infty - \infty $) in a physically meaningful way.
The modified Gottfried sum rule obtained by this method has made
clear how the correction to $1/3$ enters and its value agrees with the experimentally
determined Gottfried sum. Thus we can say that the phenomenologically determined 
quark distributions to satisfy the modified Gottfried sum rule has
taken into account the soft pion contribution effectively. Then the mean charge 
sum rule for the sea quark stands on the same theoretical footing as the modified
Gottfried sum rule. However this sum rule is badly broken in the usual
parameterization of the sea quarks. We show that the addition of the soft pion
contribution to the usual strange sea quark distribution given by the ansatz (22)
not only satisfy this sum rule but also may remove a small discrepancy in the phenomenological 
analysis. We also estimate the contribution to the structure function $g_1$ and show that
it is indispensable in the structure function in the small $x$ region. 
\\
\\
I would like to thank the warm hospitality at the CSSM in the University of Adelaide 
during the period of the ``Workshop on light-cone QCD and non-perturbative hadron
physics''(13-22,December,1999) where a part of this work is done.

\end{document}